# On the strange vector form factors of the nucleon in the NJL soliton model[†]

H. Weigel[‡], A. Abada, R. Alkofer, and H. Reinhardt

Institute for Theoretical Physics
Tübingen University
Auf der Morgenstelle 14
D-72076 Tübingen, Germany

### ABSTRACT

Within the Nambu–Jona–Lasinio model strange degrees of freedom are incorporated into the soliton picture using the collective approach of Yabu and Ando. The form factors of the nucleon associated with the nonet vector current are extracted. The numerical results provide limits for the strange magnetic moment: $-0.05 \le \mu_s \le 0.25$. For the strange magnetic form factor of the nucleon the valence quark and vacuum contributions add coherently while there are significant cancellations for the strange electric form factor.

---

[†] Supported by the Deutsche Forschungsgemeinschaft (DFG) under contract number Re–856/2-2.
[‡] Supported by a Habilitanden scholarship of the DFG.

## 1. Introduction

A number of experiments, which have recently been completed [1] or are up–coming [2, 3], measure parity violating asymmetries in scattering processes of polarized electrons on nuclei. Although these experiments were initiated as precision tests of the electro–weak theory they also provide access to hadronic "observables" like $\langle N|\bar{s}\gamma_\mu s|N\rangle$, which *e.g.* enter the matrix elements of the neutral current between nucleon states. From a theoretical point of view it is, of course, challenging to attempt predictions on these matrix elements. A first estimate of the matrix element of the strange vector current between nucleon states was carried out in ref. [4] performing a three–pole vector meson fit to dispersion relations [5]. Later on the matrix element $\langle N|\bar{s}\gamma_\mu s|N\rangle$ has been studied in the Skyrme model [6] and the Skyrme model with vector mesons [7]. Also the effect of $\phi - \omega$ mixing in the framework of vector meson dominance has been investigated [6]. This picture has even been combined [8] with the kaon loop calculation of ref. [3].

These investigations are based on chirally invariant models describing the interaction between strange mesons and the nucleon. The latter is either considered as an "elementary" particle [3, 8] or as a soliton of meson configurations [6, 7]. Neither of these studies takes explicit account of the quark structure of the nucleon. In order to examine effects related to the quark structure, the Nambu–Jona–Lasinio (NJL) model [9] represents an excellent candidate. Imitating the quark flavor dynamics of QCD at low energies the NJL model contains both reference to explicit quark degrees of freedom as well as the fruitful concepts of chiral symmetry and its spontaneous breaking. The model provides a fair description of the pseudoscalar and vector mesons as quark–antiquark bound states [10, 11, 12]. In addition, it contains soliton solutions [13] in the two flavor subspace which may be identified as baryons[a]. The so–called collective approach [15, 16] allows one to incorporate strange degrees of freedom. Its application within the NJL model [17, 18] provides an appealing possibility to explore the matrix elements of the strange vector current.

## 2. Baryons in the NJL model

The Lagrangian for the NJL model with scalar and pseudoscalar degrees of freedom is defined as the sum of the free Dirac Lagrangian and a chirally invariant four quark interaction [9]

$$\mathcal{L} = \bar{q}(i\slashed{\partial} - \hat{m}^0)q + 2G\sum_{i=0}^{N_f^2-1}\left((\bar{q}\frac{\lambda^i}{2}q)^2 + (\bar{q}\frac{\lambda^i}{2}i\gamma_5 q)^2\right). \qquad (1)$$

where $G$ denotes the effective coupling constant and $\hat{m}^0$ is the current quark mass matrix. Here we are interested in the case of three flavors, *i.e.* $N_f = 3$. By path integral bosonization the model can be converted into an effective meson theory yielding the action [10]

$$\mathcal{A} = \mathcal{A}_f + \mathcal{A}_m = \mathrm{Tr}_\Lambda\, \log\,(i\slashed{D}) - \frac{1}{4G}\mathrm{tr}\int d^4x\left(M^\dagger M - \hat{m}^0(M + M^\dagger) + (\hat{m}^0)^2\right), \qquad (2)$$

Here $M = S + iP$ contains the scalar ($S$) and pseudoscalar ($P$) meson fields. Furthermore

$$i\slashed{D} = i\slashed{\partial} - \left(P_R M + P_L M^\dagger\right) \qquad P_{R,L} = (1 \pm \gamma_5)/2. \qquad (3)$$

is the Dirac operator. We will assume isospin symmetry, $m_u^0 = m_d^0 =: m^0$; however, $m_s^0 \neq m^0$. In eq (2) we have indicated that the functional trace is UV divergent and has to be regularized.

---

[a]For a review see ref. [14] and references therein.

Hence one more parameter is introduced, the cut–off $\Lambda$. As regularization scheme we will solely employ Schwinger's proper time prescription. This requires a Wick rotation to Euclidean space ($x_0 \to -ix_4 = -i\tau$). Subsequently the real part of the Euclidean action is represented by a parameter integral

$$\frac{1}{2}\text{Tr}_\Lambda \log \left[\slashed{D}_E \slashed{D}_E^\dagger\right] = -\frac{1}{2}\text{Tr} \int_{1/\Lambda^2}^\infty \frac{ds}{s} \exp\left[-s\slashed{D}_E \slashed{D}_E^\dagger\right]. \quad (4)$$

Here $\slashed{D}_E$ refers to the continuation to Euclidean space.

The Schwinger–Dyson equations (SDEs) for the scalar fields yield non–vanishing vacuum expectation values $\langle S \rangle = \hat{m} = \text{diag}(m_u, m_d, m_s) = \text{diag}(m, m, m_s) \neq \hat{m}^0$, to represent the constituent quark masses and are a manifestation of spontaneous breaking of chiral symmetry.

In order to apply the SU(3) collective description for baryons as chiral solitons we adopt the parametrization

$$M(\mathbf{r}, t) = R(t)\xi(\mathbf{r}) R^\dagger(t) \langle S \rangle R(t) \xi(\mathbf{r}) R^\dagger(t), \quad (5)$$

for the (pseudo–)scalar fields [17]. The SU(3)–matrix $R(t)$ contains the collective coordinates. For simplicity, we have constrained the scalar fields to their vacuum configuration. The soliton configuration is characterized by the hedgehog *ansatz* for the chiral field

$$U(\mathbf{r}) = \xi^2(\mathbf{r}) = \exp\left(i\hat{\mathbf{r}} \cdot \boldsymbol{\tau}\Theta(r)\right). \quad (6)$$

The special form of the parametrization (5) guarantees that only the hedgehog rotates in the space of the collective coordinates. It is advantageous to transform to the flavor rotating frame $q' = R^\dagger(t)q$, *i.e.*

$$i\beta R^\dagger \slashed{D}_E R = i\beta \slashed{D}_E' = -\partial_\tau - h_\Theta - h_{\text{rot}} - h_{\text{SB}} \quad (7)$$

Here $h_\Theta = \boldsymbol{\alpha}\cdot\mathbf{p} + \beta m U^{\gamma_5}$ refers to the one–particle Dirac Hamiltonian in the background of the chiral soliton. Furthermore

$$h_{\text{rot}} = \frac{i}{2}\lambda^a \Omega_E^a \quad (8)$$

contains the analytic continuation ($\Omega_E^a = -i\Omega^a$) of the angular velocity measuring the time dependence of the collective coordinates, $R^\dagger \dot{R} = (i/2)\lambda^a \Omega^a$. The symmetry breaking piece is linear in the difference of the constituent quark masses[b]

$$h_{\text{SB}} = \frac{m - m_s}{\sqrt{3}} \mathcal{T} \beta \left(\sum_{a=1}^{8} D_{8a}\lambda^a - \lambda^8\right) \mathcal{T}^\dagger. \quad (9)$$

The static soliton enters $h_{\text{SB}}$ via $\mathcal{T} = \xi P_L + \xi^\dagger P_R$. Furthermore the adjoint representation of the collective rotations $D_{ab} = (1/2)\text{tr}(\lambda^a R \lambda^b R^\dagger)$ has been introduced.

Then $\mathcal{A}$ is expanded up to quadratic order in terms of the angular velocity as well as the mass difference $m - m_s$. From the resulting expression the collective Lagrangian $L = L(R, \Omega^a)$ is straightforwardly extracted[17]. Quantization is achieved by identifying the momenta conjugate to the angular velocities with the right generators of SU(3)

$$\mathcal{R}_a = -\frac{\partial L(R, \Omega^a)}{\partial \Omega_a}, \quad (10)$$

---

[b]The explicit dependence on the current quark mass difference $m_u^0 - m_s^0$ is completely contained in $\mathcal{A}_m$.

which provides a linear relation between the generators and the velocities. The resulting collective Hamiltonian $H = H(R, \mathcal{R}_a)$ can be diagonalized exactly. For details we refer to the literature [16, 6, 17]. Here we only wish to mention that due to SU(3) symmetry breaking the eigenfunctions of $H$ are distorted SU(3) D–functions.

## 3. Vector currents in the SU(3) NJL model

The vector currents $J^a_\mu$ are most easily obtained by introducing external sources $b^a_\mu$, which couple to the quark bilinear $\bar{q}\gamma_\mu(\lambda^a/2)q$. Then $J^a_\mu$ is the derivative of the extended action with respect to these sources

$$J^a_\mu(x) = \frac{-i\delta}{\delta b^{\mu a}(x)} \mathrm{Tr} \log\left[i\beta \not{D}\left(b^a_\mu\right)\right]\Big|_{b^a_\mu=0} . \tag{11}$$

Transforming to the flavor rotating frame and performing the Wick rotation the extended Dirac operator reads

$$i\beta \not{D}'_E\left(b^a_\mu\right) = -\partial_\tau - h_\Theta - h_{\mathrm{rot}} - h_{\mathrm{SB}} - i b^a_4 D_{ab}\frac{\lambda^b}{2} - \boldsymbol{\alpha} \cdot \boldsymbol{b}^a D_{ab}\frac{\lambda^b}{2}, \tag{12}$$

where $b^a_4 = ib^a_0$ denotes the analytic continuation of the time component of the source current. $\lambda^a$ labels the generators of the flavor group, including the singlet piece $\lambda^0 = \sqrt{2/3}\,\mathbb{1}$. The latter is of special importance for the computation of the strange vector current because it enters the projector onto strange degrees of freedom, $\mathrm{diag}(0,0,1) = \lambda^0/\sqrt{6} - \lambda^8/\sqrt{3}$.

To obtain a normalization consistent with (10) of the symmetry charges the currents have to be expanded up to linear order in both $h_{\mathrm{rot}}$ and $h_{\mathrm{SB}}$. Induced kaon components have to be included in a way consistent with the computation of the strange moment of inertia $\beta^2$ [17]. Leaving symmetry breaking effects apart, the time independence of the charges associated with the vector current requires to consider $h_{\mathrm{rot}}$ (8) (and therefore $\Omega^a$) as a classical quantity constant in time.

The currents gain contributions from both, the explicit occupation of the valence quark orbits and the polarized vacuum [19]. The former is obtained by a perturbation expansion for the valence quark level[c] $\Psi_{\mathrm{val}}$, i.e. by substituting

$$\Psi = R\left\{\Psi_{\mathrm{val}} + \sum_{\mu \ne \mathrm{val}} \Psi_\mu \frac{\langle\mu|h_{\mathrm{rot}} + h_{\mathrm{SB}}|\mathrm{val}\rangle}{\epsilon_{\mathrm{val}} - \epsilon_\mu}\right\} \tag{13}$$

into the expression for the vector current $\bar{\Psi}\gamma_\mu(\lambda^a/2)\Psi$ yielding

$$J^{a(\mathrm{val})}_\mu = \frac{N_C}{2}D_{ab}\bar{\Psi}_{\mathrm{val}}\gamma_\mu\lambda^b\Psi_{\mathrm{val}} + \frac{N_C}{2}D_{ab}\sum_{\nu \ne \mathrm{val}}\left\{\bar{\Psi}_\nu\gamma_\mu\lambda^b\Psi_{\mathrm{val}}\frac{\langle\mathrm{val}|\Omega^c\frac{\lambda^c}{2} + h_{\mathrm{SB}}|\nu\rangle}{\epsilon_{\mathrm{val}} - \epsilon_\nu} + \mathrm{h.\,c.}\right\}. \tag{14}$$

In (13) and (14) $\epsilon_\mu$ and $\Psi_\mu$ refer to the eigenvalues and eigenstates of $h_\Theta$.

The functional traces are evaluated imposing anti–periodic boundary conditions for the quark fields in a Euclidean time interval $T$, with the vacuum contribution obtained from (11) in the limit $T \to \infty$. The relevant techniques may be found in refs. [19, 17]. We thus only quote the final result for the expansion of (11) up to linear order in $\Omega^a$ and the mass difference $m - m_s$ (contained in $h_{\mathrm{SB}}$)

$$\begin{aligned}J^{a(\mathrm{vac})}_\mu &= \frac{N_C}{4}D_{ab}\sum_\nu \bar{\Psi}_\nu\gamma_\mu\lambda^b\Psi_\nu\,\mathrm{sgn}\left(\epsilon_\nu\right)\mathrm{erfc}\left(\left|\frac{\epsilon_\nu}{\Lambda}\right|\right) \\ &\quad + \frac{N_C}{4}D_{ab}\Omega^c\sum_{\nu\rho}\bar{\Psi}_\nu\gamma_\mu\lambda^b\Psi_\rho\langle\rho|\lambda^c\Omega^c f^{\nu\rho}_{\mathrm{rot}} + h_{\mathrm{SB}} f^{\nu\rho}_{\mathrm{SB}}|\nu\rangle\end{aligned} \tag{15}$$

---

[c]The valence quark level is defined as the state with the eigenvalue of smallest module.

with the regularization functions

$$f_{\text{SB}}^{\nu\rho} = \frac{\text{sgn}(\epsilon_\nu)\,\text{erfc}\left(\left|\frac{\epsilon_\nu}{\Lambda}\right|\right) - \text{sgn}(\epsilon_\rho)\,\text{erfc}\left(\left|\frac{\epsilon_\rho}{\Lambda}\right|\right)}{\epsilon_\nu - \epsilon_\rho}, \quad f_{\text{rot}}^{\nu\rho} = \frac{1}{2} f_{\text{SB}}^{\nu\rho} - \frac{\Lambda}{\sqrt{\pi}} \frac{e^{-\epsilon_\rho^2/\Lambda^2} - e^{-\epsilon_\nu^2/\Lambda^2}}{\epsilon_\rho^2 - \epsilon_\nu^2}. \quad (16)$$

In (15) we have chosen to regularize the UV–finite imaginary part of the Euclidean action as well. Omitting this regularization, which essentially corresponds to a different model, amounts to assuming the limit $\Lambda \to \infty$ in those terms which are related to an odd number of time components of Lorentz vectors. In this context $\Omega^c$ has to be counted as a time component. The total current is the sum

$$J_\mu^a = \eta_{\text{val}} J_\mu^{a(\text{val})} + J_\mu^{a(\text{vac})} \quad (17)$$

with $\eta_{\text{val}} = 0, 1$ adjusted to describe a unit baryon number configuration. From eqs (14,15) it is suggestive that the current may be cast into a sum of products of radial functions $V_l(r)$ and isospin covariant expressions involving the SU(3) collective coordinates $D_{ab}$ and the velocities $\Omega^a$ [7]. Upon (10) the latter are replaced by the SU(3) generators $\mathcal{R}_a$. This permits to compute the spin and flavor parts of the matrix elements of $J_\mu^a$ between baryon eigenstates[d]. We may formally write [7]

$$J_i^a = \sum_{l=1}^{6} V_l(r) \sum_{j,k=1}^{3} \epsilon_{ijk} x_j \mathcal{M}_l^{ak} \quad \text{and} \quad J_0^a = \sum_{l=9}^{13} V_l(r) \mathcal{M}_l^a. \quad (18)$$

According to (17) the radial functions $V_l(r)$ separate into valence and vacuum parts. For the explicit form of the matrix elements $\mathcal{M}$ in terms of SU(3) "Euler–angles" and their computation we refer to appendix A of ref. [6]. Furthermore we define Fourier transforms of the radial functions

$$\begin{aligned}
\tilde{V}_l(|\mathbf{q}|) &= 4\pi \int dr\, r^2 \frac{r}{|\mathbf{q}|} j_1(|\mathbf{q}|r) V_l(r)\,, \quad l = 1, ..., 6 \\
\tilde{V}_l(|\mathbf{q}|) &= 4\pi \int dr\, r^2 j_0(|\mathbf{q}|r) V_l(r)\,, \quad l = 9, ..., 13
\end{aligned} \quad (19)$$

to evaluate the spatial parts of baryon matrix elements. Identifying finally the momentum transfer in the Breit frame ($Q^2 = -\mathbf{q}^2$) allows us to extract the electric $G_E$ and magnetic $G_M$ form factors [20]

$$G_M^a(Q^2) = -2M_N \sum_{l=1}^{6} \tilde{V}_l(|\mathbf{q}|) \mathcal{M}_l^{a3} \quad \text{and} \quad G_E^a(Q^2) = \sum_{l=9}^{13} \tilde{V}_l(|\mathbf{q}|) \mathcal{M}_l^a. \quad (20)$$

The nucleon mass $M_N = 940\text{MeV}$ has been introduced because $G_M$ is measured in nucleon magnetons. It should be noted that recoil corrections have been ignored. Thus (20) is only valid in the vicinity of $|\mathbf{q}| = 0$. In (20) we have kept the full flavor structure. E.g. the electro–magnetic form factor is given by the linear combination $G_{E,M}^{e.m.} = G_{E,M}^3 + G_{E,M}^8/\sqrt{3}$ while the strange vector form factors include the singlet piece $G_{E,M}^s = G_{E,M}^0/\sqrt{6} - G_{E,M}^8/\sqrt{3}$. The magnetic moments of the nucleon are defined as the matrix elements $\mu_{p,n} = \langle \pm G_M^3(0) + G_M^8(0)/\sqrt{3}\rangle_p$ and $\mu_S = \langle G_M^0(0)/\sqrt{6} - G_M^8(0)/\sqrt{3}\rangle_p$.

---

[d]A typical example for these matrix elements is $\mathcal{M}_3^{ak} = \langle \sum_{\alpha,\beta=4}^{7} d_{k\alpha\beta} D_{a\alpha} \mathcal{R}_\beta \rangle$, where the $d_{abc}$ refer to the symmetric structure coefficients of SU(3).

Table 1: Baryon properties as a functions of the scaling variable (21). $\mu_p$ and $\mu_n$ are the magnetic moments of the proton and neutron, respectively. $\mu_S$ denotes our prediction for the strange magnetic moment of the nucleon. $\chi = [\sum_{\text{baryons}}(\triangle M_{\text{pred}} - \triangle M_{\text{expt}})^2]^{1/2}$ measures the deviation of the predicted mass differences from the experimental ones.

|  | $m$=400MeV | | | | $m$=450MeV | | | | expt. |
| --- | --- | --- | --- | --- | --- | --- | --- | --- | --- |
| $\lambda$ | 1.0 | 0.9 | 0.8 | 0.7 | 1.0 | 0.9 | 0.8 | 0.7 | |
| $\mu_p$ | 1.17 | 1.32 | 1.58 | 2.02 | 1.06 | 1.21 | 1.46 | 1.90 | 2.79 |
| $\mu_n$ | -0.76 | -0.90 | -1.17 | -1.63 | -0.69 | -0.84 | -1.09 | -1.56 | -1.91 |
| $\mu_S$ | 0.24 | 0.21 | 0.18 | 0.15 | 0.23 | 0.20 | 0.18 | 0.16 | ? |
| $\chi$(MeV) | 319 | 219 | 88 | 199 | 341 | 255 | 125 | 144 | 0 |

## 4. Numerical Results and discussion

Expanding the action up to quadratic order in the pseudoscalar fields allows one to express the corresponding masses and decay constants in terms of the parameters $\hat{m}, \hat{m}^0, G$ and $\Lambda$. Using the pion decay constant $f_\pi$=93MeV and mass $m_\pi = 135$MeV together with the SDE in the non–strange sector determines $m^0, G$ and $\Lambda$ for a given value of $m$. Then the kaon mass $m_K = 495$MeV and the SDE in the strange sector yield $m_s$ and $m_s^0$ for the same value of $m$. The kaon decay constant $f_K$, left as a prediction, is commonly underestimated in the NJL model [17]. In the baryon sector the mass differences[e] of the low–lying $\frac{1}{2}^+$ and $\frac{3}{2}^+$ baryons are found to be on the small side [17, 18]. This can be understood as being caused by the too small spatial extension of the self–consistent soliton. Denoting the self–consistent profile, which extremizes the static energy functional $E_{\text{cl}}$, by $\Theta_{\text{s.c.}}(r)$ we consider

$$\Theta_\lambda(r) = \Theta_{\text{s.c.}}(\lambda r). \qquad (21)$$

The driving symmetry breaking term entering the collective Hamiltonian stems from $\mathcal{A}_m$ and can easily be verified to be proportional to $\lambda^{-3}$. It should be remarked that $E_{\text{cl}}$ is rather insensitive to $\lambda$, choosing e.g. $0.7 \le \lambda \le 0.8$ yields an increase of only 10%. Furthermore such values for $\lambda$ are suggested by the proper inclusion of the $\omega$ meson [22] and the corresponding mass differences are in reasonable agreement with the experimental data[f]. The dependence of the magnetic moments on $\lambda$ is displayed in table 1. The isovector part of the magnetic moment of the nucleon $\mu_V = \mu_p - \mu_n$ is sensitive to the extension of the soliton configuration. For the self–consistent soliton configuration ($\lambda = 1$) $\mu_V$ is predicted to be less than half of its experimental value, 4.70. However, $\mu_V$ increases drastically with the extension of the meson profile. On the other hand the isoscalar part ($\mu_p + \mu_n$) is quite insensitive to the extension of the soliton. Then $0.7 \le \lambda \le 0.8$ provides reasonable descriptions of the baryon mass differences and the magnetic moments associated with the electro–magnetic current. Hence the results shown in table 1 suggest the prediction $0.15 \le \mu_S \le 0.2$.

Recently the discussion of $1/N_C$ corrections to $\mu_V$ [24] came into fashion[g]. As these have the same dependence on the collective coordinates their incorporation is effectively equivalent to a modification of $\tilde{V}_i(|\boldsymbol{q}| = 0)$. Actually these corrections are relevant only for the radial function already present in the two flavor model ($V_1$ in the notation of ref. [7]). Without going into detail one may therefore adjust $\tilde{V}_1(|\boldsymbol{q}|=0)$ to reproduce $\mu_V$. Since this represents quite a sizable change it provides an upper bound for the $1/N_C$ corrections to the magnetic form factors. Keeping all other quantities at the values associated with the self–consistent soliton the strange magnetic moment of the nucleon becomes $\mu_S = -0.03(-0.05)$ for $m = 400(450)$MeV. Apparently modifications, which lead to a larger $\mu_V$ causes $\mu_S$ to decrease. From this we consider the result $\mu_S \approx 0.25$ obtained from the self–consistent soliton as an upper bound. $\mu_S$ may even assume small negative values as in the Skyrme model with vector mesons [7].

---

[e] For the description of the absolute values of the baryon masses quantum corrections due to meson fluctuations have to be taken into account [21].

[f] Considering $\lambda$ as a variable quantity is furthermore motivated by the existence of sizable correlations between SU(3) symmetry breaking and the extension of the soliton [23].

[g] Here we will not pursue the question of validity of the treatments in ref. [25].

Table 2: Electric radii of the nucleon as functions of the scaling variable (21). Data are in fm$^2$.

|  | $m=400$MeV | | | | $m=450$MeV | | | | expt. |
| --- | --- | --- | --- | --- | --- | --- | --- | --- | --- |
| $\lambda$ | 1.0 | 0.9 | 0.8 | 0.7 | 1.0 | 0.9 | 0.8 | 0.7 |  |
| $\langle r^2 \rangle_p$ | 0.66 | 0.81 | 1.06 | 1.43 | 0.61 | 0.78 | 1.03 | 1.40 | 0.74 |
| $\langle r^2 \rangle_n$ | -0.18 | -0.21 | -0.28 | -0.43 | -0.16 | -0.20 | -0.28 | -0.43 | -0.12 |
| $\langle r^2 \rangle_S$ | 0.03 | -0.08 | -0.22 | -0.39 | -0.03 | -0.12 | -0.26 | -0.41 | ? |

The electric radii of the nucleon, which are given by the slopes of the associated electric form factors $\langle r^2 \rangle = -6\partial G_E(Q^2)/\partial Q^2|_{Q^2=0}$ are shown in table 2. Not surprisingly the radii increase with decreasing $\lambda$ and the experimental data are reasonably reproduced for a somewhat larger value of the scaling variable than in the case of the magnetic moments. Within the reliability of the approach we estimate $\langle r_S^2 \rangle \approx -0.1... -0.2 \text{fm}^2$.

Figure 1 exhibits that the valence quark and vacuum contributions to $G_{E,S}$ of the nucleon almost cancel each other leading to a total strange electric form factor which barely deviates from zero. The decomposition into $G_S^{\text{val}}$ and $G_S^{\text{vac}}$ refers to (17). On the other hand $G_{M,S}^{\text{val}}$ and $G_{M,S}^{\text{vac}}$ add up coherently to $G_{M,S}$. While $G_{M,S}^{\text{val}}$ drops monotonously the vacuum part develops an extremum which is also recognized in $G_{M,S}$. A non-monotonous behavior for $G_{M,S}$ is also known from the Skyrme model with vector meson [7].

We have also considered the model with the imaginary part unregularized. For the self–consistent soliton this yields a slight increase for the strange magnetic moment $\mu_S \approx 0.35$ while $\langle r_S^2 \rangle$ remains within the bounds given above.

## 5. Summary

Using the generalized Yabu–Ando approach to the chiral soliton of the NJL model we have investigated the matrix elements of the nonet vector current between nucleon states. These are of great interest because they provide information on the strangeness content of the nucleon. Furthermore these matrix elements enter the analysis of experiments attempting precision tests of the standard model.

To some extend the model is plagued by the too small prediction for the isovector magnetic moment, $\mu_V$, of the nucleon. Modifications improving on this prediction cause the strange magnetic moment, $\mu_S$, to decrease. As a consequence the result related to the self–consistent soliton represents the upper bound. A lower bound has been obtained by (over) estimating possible $1/N_C$ effects, yielding

$$-0.05 \leq \mu_S \leq 0.25. \tag{22}$$

The lower bound is compatible with the corresponding prediction in the Skyrme model with vector mesons [7]. Other models give even smaller results [8]. We have furthermore observed that the valence quark contribution (*i.e.* short range effects) dominates $\mu_S$ although the vacuum part adds coherently to the strange magnetic form factor. This is in contrast to the strange electric form factor which is characterized by a large cancellation between the valence quark and vacuum contributions and is approximately zero. The NJL model prediction for the slope of this form factor is

$$-0.2\text{fm}^2 \leq \langle r_S^2 \rangle \leq -0.1\text{fm}^2. \tag{23}$$

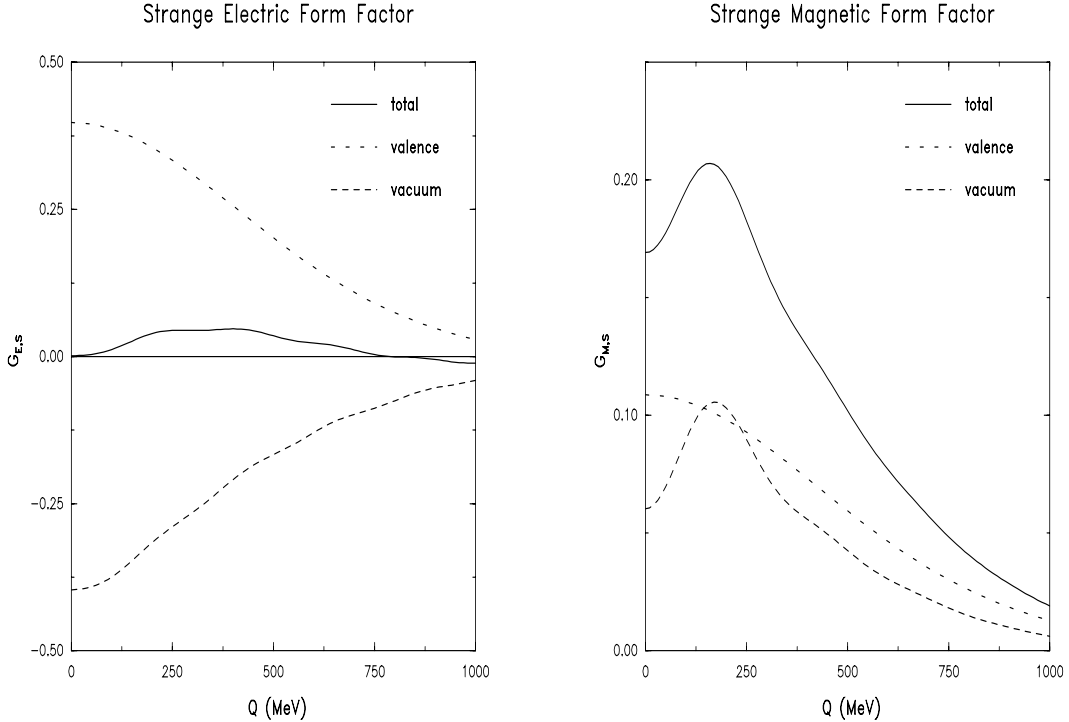

Figure 1: The strange electric and magnetic form factors of the nucleon. Parameters are $m = 450\text{MeV}$ and $\lambda = 0.75$, which provide a good agreement for the baryon mass differences: $\chi = 68\text{MeV}$, *cf.* table 1.

This result compares well with that of the Skyrme model [6]. As $G_{E,S} \approx 0$ it is easy to understand that small effects may modify this result. As an example we refer to the $\phi - \omega$ mixing, which is neither included in the present model nor in the Skyrme model.

*References*

[1] P. Souder et al., Phys. Rev. Lett. **65** (1990) 694;
W. Heil et al., Nucl. Phys. **B237** (1989) 1.

[2] J. Ahrens et al., Nucl. Phys. News **4** (1994) 5.

[3] M. J. Musolf and T. W. Donnelly, Z. Phys. **C57** (1993) 559.

[4] R. Jaffe, Phys. Lett. **B229** (1989) 275.

[5] G. Höhler et al., Nucl. Phys. **B114** (1974) 505.

[6] N. W. Park, J. Schechter, and H. Weigel, Phys. Rev. **D43** (1991) 869.

[7] N. W. Park and H. Weigel, Nucl. Phys. **A541** (1992) 453.

[8] T. Cohen, H. Forkel, and M. Nielsen, Phys. Lett. **B316** (1993) 1.
H. Forkel, M. Nielsen, X.-M. Jin, and T. Cohen, *Stranger in the light ...*, Maryland University preprint, hep-ph/9408326, March 1994.

[9] Y. Nambu and G. Jona-Lasinio, Phys. Rev. **122** (1961) 345; **124** (1961) 246.

[10] D. Ebert and H. Reinhardt, Nucl. Phys. **B271** (1986) 188.

[11] V. Bernard and U.–G. Meißner, Nucl. Phys. **A489** (1988) 647.

[12] T. Hatsuda and T. Kunihiro, Phys. Rep. **247** (1994) 221.

[13] H. Reinhardt and R. Wünsch, Phys. Lett. **215** (1988) 577; **B 230** (1989) 93;
T. Meißner, F. Grümmer, and K. Goeke, Phys. Lett. **B 227** (1989) 296;
R. Alkofer, Phys. Lett. **B 236** (1990) 310.

[14] R. Alkofer, H. Reinhardt, and H. Weigel, *Baryons as Chiral Solitons in the Nambu–Jona-Lasinio Model*, Tübingen University preprint, UNITU-THEP-25/1994, hep-ph/9501213.

[15] G. S. Adkins, C. R. Nappi, and E. Witten, Nucl. Phys. **B228** (1983) 552;
E. Guadagnini, Nucl. Phys. **B236** (1984) 15.

[16] H. Yabu and K. Ando, Nucl. Phys. **B301** (1988) 601.

[17] H. Weigel, R. Alkofer, and H. Reinhardt, Nucl. Phys. **B387** (1992) 638.

[18] A. Blotz, D. Diakonov, K. Goeke, N. W. Park, V. Petrov, and P. V. Pobilitsa, Nucl. Phys. **A555** (1993) 765.

[19] H. Reinhardt, Nucl. Phys. **A503** (1989) 825.

[20] E. Braaten, S.–M. Tse and C. Willcox, Phys. Rev. Lett. **56** (1986) 2008; Phys. Rev. **D34** (1986) 1482;
U.-G. Meißner, N. Kaiser, and W. Weise, Nucl. Phys. **A466** (1987) 685.

[21] G. Holzwarth, Nucl. Phys. **A572** (1994) 69;
H. Weigel, R. Alkofer, and H. Reinhardt, Nucl. Phys. **A582** (1995) 484.

[22] H. Weigel, U. Zückert, R. Alkofer, and H. Reinhardt, *On the analytic properties of chiral solitons in the presence of the $\omega$ meson*, Tübingen University preprint, hep-ph/9407304, July 1994, Nucl. Phys. **A**, in press.

[23] J. Schechter and H. Weigel, Phys. Rev. **D44** (1991) 2916.

[24] C. R. Dashen, E. Jenkins, and A. V. Manohar, Phys. Rev. **D49** (1994) 4713.

[25] M. Wakamatsu and T. Watabe, Phys. Lett. **B312** (1993) 184;
A. Blotz, M. Prasałowicz and K. Goeke, Phys. Lett. **B317** (1993) 195;
R. Alkofer and H. Weigel, Phys. Lett. **B319** (1993) 1;
C. V. Christov, K. Goeke, P. Pobilitsa, V. Petrov, M. Wakamatsu, and T. Watabe, Phys. Lett. **B325** (1994) 467.